\documentclass{PoS}

\usepackage{graphicx}
\usepackage{amsmath}
\usepackage{amssymb}

\newcommand{\be}{\begin{equation}}
\newcommand{\ee}{\end{equation}}
\newcommand{\bea}{\begin{eqnarray}}
\newcommand{\eea}{\end{eqnarray}}
\newcommand{\bi}{\begin{itemize}}
\newcommand{\ei}{\end{itemize}}
\newcommand{\ben}{\begin{enumerate}}
\newcommand{\een}{\end{enumerate}}
\newcommand{\bt}{\begin{tabbing}}
\newcommand{\et}{\end{tabbing}}

\newcommand{\ff}{F_V^{P}}
\newcommand{\fflo}{F_{V,0}^P}
\newcommand{\ffnlo}{F_{V,2}^P}
\newcommand{\ffnnlo}{F_{V,4}^P}
\newcommand{\ffnnnlo}{F_{V,6}^P}
\newcommand{\ffnloL}{F_{V,2,L}^P}
\newcommand{\ffnloB}{F_{V,2,B}^P}
\newcommand{\ffnnloL}{F_{V,4,L}^P}
\newcommand{\ffnnloC}{F_{V,4,C}^P}
\newcommand{\ffnnloB}{F_{V,4,B}^P}
\newcommand{\pffnloL}{F_{V,2,L}^{\pi^+}}

\newcommand{\pffnnloC}{F_{V,4,C}^{\pi^+}}

\newcommand{\kpff}{F_{V}^{K^+}}

\newcommand{\kpffnlo}{F_{V,2}^{K^+}}
\newcommand{\kpffnnlo}{F_{V,4}^{K^+}}
\newcommand{\kpffnloL}{F_{V,2,L}^{K^+}}

\newcommand{\kpffnnloC}{F_{V,4,C}^{K^+}}

\newcommand{\knff}{F_V^{K^0}}

\newcommand{\knffnloL}{F_{V,2,L}^{K^0}}

\newcommand{\knffnnloC}{F_{V,4,C}^{K^0}}

\newcommand{\pkpff}{F_{V}^{\{\pi^+,K^+\}}}

\newcommand{\cradpff}{{\langle r^2 \rangle_V^{\pi^+}}}

\newcommand{\cradkpff}{{\langle r^2 \rangle_V^{K^+}}}

\newcommand{\cradknff}{{\langle r^2 \rangle_V^{K^0}}}

\newcommand{\Cppt}{{c_{\pi^+,\pi t}^r}}
\newcommand{\Cpkt}{{c_{\pi^+,K t}^r}}
\newcommand{\Ctt}{{c_{t^2}^r}}
\newcommand{\Ckpt}{{c_{K^+,\pi t}^r}}
\newcommand{\Ckkt}{c_{K^+,K t}^r}
\newcommand{\Ckn}{{c_{K^0}^r}}

\newcommand{\fzt}{{\tilde{f}_0}}
\newcommand{\Cplspk}{{c_{+,\pi K}^r}}

\title{
   \begin{picture}(0,0)(0,0)%
   \put(350,75){\makebox(0,0)[l]{\textnormal{\normalsize KEK-CP-329}}}%
   \end{picture}%
   Chiral behavior of light meson form factors 
   in 2+1 flavor QCD with exact chiral symmetry
}

\ShortTitle{
   Chiral behavior of light meson form factors 
   in 2+1 flavor QCD with exact chiral symmetry
}

\author{
   JLQCD Collaboration: 
   \speaker{T.~Kaneko}$^{a,b}$\thanks{E-mail: takashi.kaneko@kek.jp},
   S.~Aoki$^{c,d}$, 
   G.~Cossu$^a$, 
   X.~Feng$^e$, 
   H.~Fukaya$^f$, 
   S.~Hashimoto$^{a,b}$,  
   J.~Noaki$^a$, 
   T.Onogi$^f$
   \\
   \\
   \\
   \llap{$^a$}
   High Energy Accelerator Research Organization (KEK),
   Ibaraki 305-0801, Japan 
   \\
   \llap{$^b$}
   School of High Energy Accelerator Science,
   SOKENDAI (The Graduate University for Advanced Studies),
   Ibaraki 305-0801, Japan
   \\
   \llap{$^c$}
   Yukawa Institute for Theoretical Physics,
   Kyoto University, 
   Kyoto 606-8502, Japan
   \\
   \llap{$^d$}
   Center for Computational Sciences, University of Tsukuba, 
   Ibaraki 305-8577, Japan  
   \\
   \llap{$^e$}
   Physics Department, Columbia University, 
   New York, NY 10027, USA
   \\
   \llap{$^f$}
   Department of Physics, Osaka University, 
   Osaka 560-0043, Japan
}

\abstract{
We present a study of chiral behavior of light meson 
form factors in QCD with three flavors of overlap quarks. 
Gauge ensembles are generated at single lattice spacing 0.12\,fm
with pion masses down to 300 MeV. The pion and kaon 
electromagnetic form factors and the kaon semileptonic 
form factors are precisely calculated using the all-to-all 
quark propagator. We discuss their chiral behavior 
using the next-to-next-to-leading order chiral perturbation theory.
}

\FullConference{The 33rd International Symposium on Lattice Field Theory\\
		14 -18 July 2015\\
		Kobe International Conference Center, Kobe, Japan*}

\begin{document}


\section{Introduction}

The $K \!\to\! \pi l \nu$ semileptonic decays provide
a precise determination of the Cabibbo-Kobayashi-Maskawa matrix element 
$|V_{us}|$.
Lattice QCD plays an important role to evaluate 
the normalization of the vector and scalar form factors $f_+(0)\!=\!f_0(0)$ 
in the $K\!\to\!\pi$ matrix element
\bea
   \langle \pi(p^\prime) | V_\mu | K(p) \rangle 
   & = & 
   (p+p^\prime)_\mu f_+(t) 
 + \frac{M_K^2-M_\pi^2}{t}\, q_\mu \left\{ f_0(t)-f_+(t) \right\}
   \hspace{5mm}
   \left(t=q^2=(p-p^\prime)^2\right).
   \hspace{8mm}
   \label{eqn:intro:ME}
\eea
The required accuracy is high, typically within 1\%.
The phase space integral for the decay rate is estimated from the 
form factor shape, namely their $t$ dependence, which has been precisely 
measured by experiments. Therefore a rigorous comparison of 
the shape between lattice QCD and experiments can demonstrate 
the reliability of the precision calculation of $f_+(0)$.
This article presents 
a lattice calculation of these form factors and analysis 
based on next-to-next-to-leading order (NNLO) chiral perturbation theory (ChPT).
We also discuss the kaon and pion electromagnetic (EM) form factors
defined through 
\bea
   \langle P(p^\prime) | J_\mu | P(p) \rangle 
   & = &
   \left( p + p^\prime \right)_\mu \ff(t)
   \hspace{5mm}
   (P = \pi^+,K^+,K^0),
   \label{eqn:intro:ff}
\eea
which provide helpful information for the ChPT analysis of the 
semileptonic form factors.


\section{Simulation method}


We simulate $2+1$ flavor QCD using the overlap quark action, 
which exactly preserves chiral symmetry and enables us to 
directly compare the lattice data with ChPT.
Numerical simulations are remarkably accelerated 
by modifying the Iwasaki gauge action~\cite{exW}
and by simulating the trivial topological sector~\cite{exW,fixedQ}.
Note that effects of the fixed global topology 
can be considered as finite volume effects 
suppressed by the inverse lattice volume~\cite{fixedQ}. 
Gauge ensembles are generated at a lattice spacing $a\!=\!0.112(1)$~fm
and at a strange quark mass $m_s\!=\!0.080$
close to its physical value $m_{s,\rm phys}\!=\!0.081$.
Four values of degenerate up and down quark masses,
$m_l\!=\!0.015$, 0.025, 0.035 and 0.050, are simulated 
to explore a range of the pion mass 290\,--\,540~MeV.
At each $m_l$, 
we choose a lattice size, $16^3 \!\times\! 48$ or $24^3 \!\times\! 48$,
to control finite volume effects
by satisfying a condition $M_\pi L \! \gtrsim \! 4$.
The statistics are 2,500 HMC trajectories 
at each simulation point $(m_l,m_s)$.


We calculate the relevant two- and three-point functions of 
pion and kaon by using the all-to-all quark propagator. 
The form factors are precisely estimated from ratios 
of the correlation functions. 
We employ the twisted boundary condition for the valence quarks
to simulate near-zero momentum transfers $|t|\!\lesssim\!(300~\mbox{MeV})^2$.
The $m_s$ dependence of the form factors is studied 
by repeating our calculation at a different $m_s$ ($=\!0.060$)
with the reweighting technique. 
We refer to Refs~\cite{Kl3:Nf3:JLQCD:Lat12,EMFF:Nf3:JLQCD}
for more details on our simulation parameters and method.


\section{EM form factors}

\begin{figure}[t]
\begin{center}
   \includegraphics[angle=0,width=0.48\linewidth,clip]%
                   {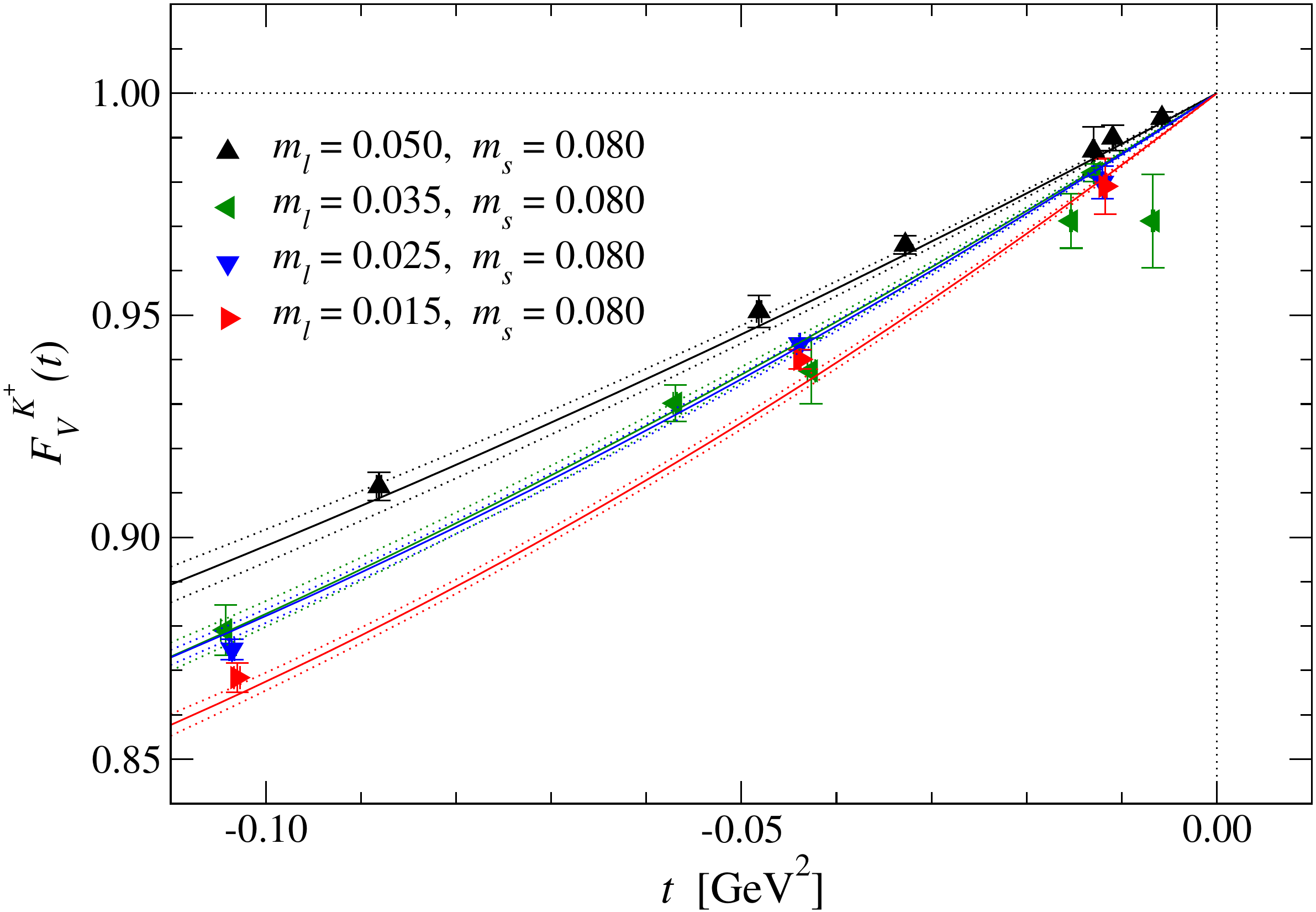}
   \hspace{3mm}
   \includegraphics[angle=0,width=0.48\linewidth,clip]%
                   {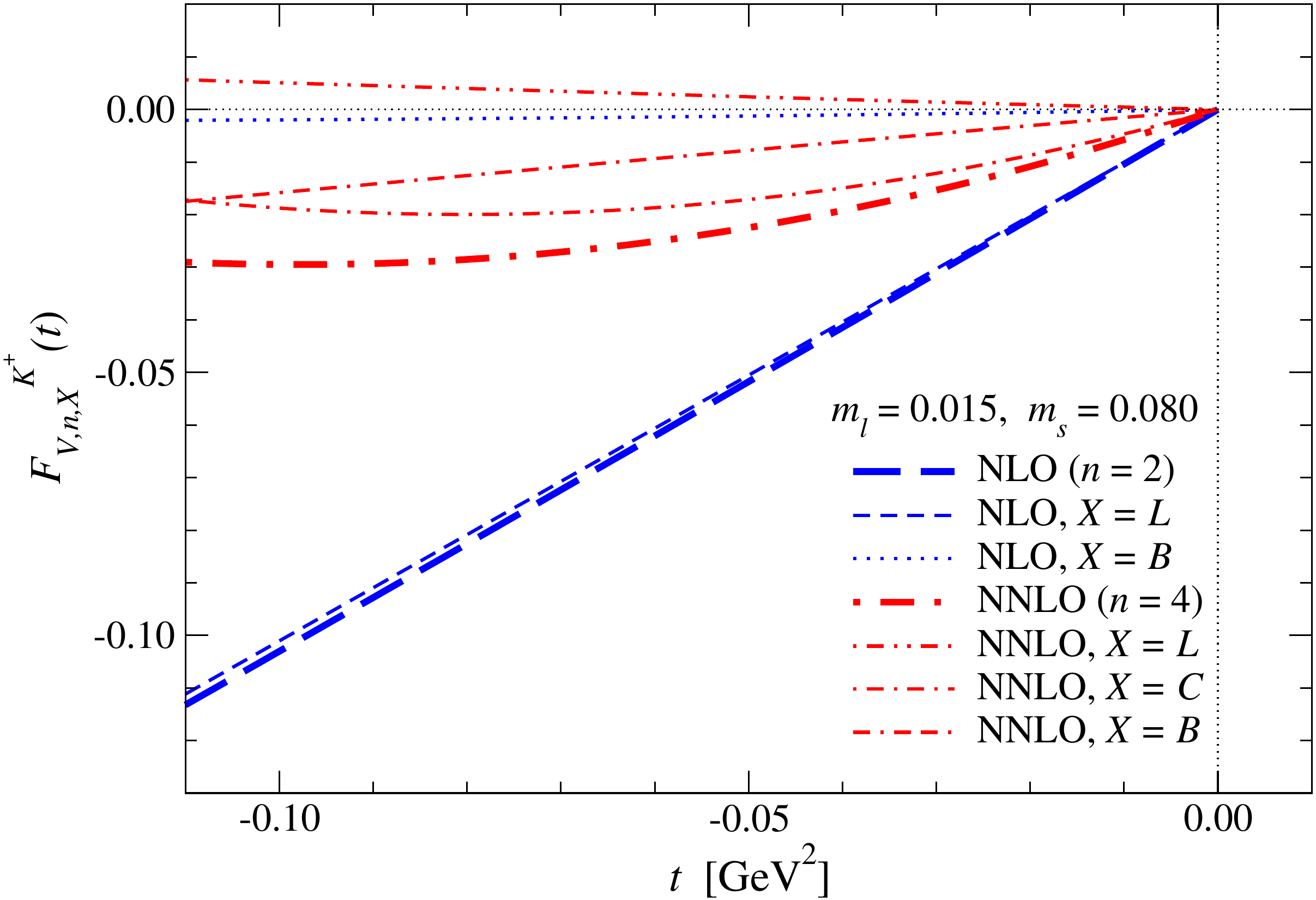}
   \vspace{-1mm}
   \caption{
     Left panel: 
     NNLO ChPT fit to $K^+$ EM form factor.
     Data at different $M_\pi$'s are plotted in different symbols 
     as a function of $t$.
     Right panel: 
     NLO and NNLO contributions (thick lines) and 
     their decomposition into LEC-dependent and independent parts
     (thin lines)
     at $(m_l,m_s)\!=\!(0.015,0.080)$. 
     The blue and red lines show the NLO and NNLO contributions,
     respectively.
   }
   \label{fig:em}
\end{center}
\vspace{-5mm}
\end{figure}

Chiral symmetry constrains the chiral behavior of the EM form factors.
In the chiral expansion of these form factors 
at the next-to-leading order (NLO) 
in terms of the expansion parameters 
$\xi_{\{\pi,K\}}\!=\!M_{\{\pi,K\}}^2/(4\pi F_\pi)^2$ and $t$,
we have only single free parameter $L_9^r$,
which is one of the low-energy constants (LECs) 
in the NLO chiral Lagrangian.
Many more LECs appear at NNLO: 
$L_{\{1,...,5\}}^r$ and $C_i^r$'s in the NLO and NNLO Lagrangians.
Since $L_{\{1,...,5\}}^r$ have been well studied in phenomenological 
studies of experimental data and they appear only in the 
possibly small NNLO corrections,
we fix them to a recent phenomenological estimate 
in Ref.~\cite{ChPT:LECs:SU2+SU3}. 
In general, $C_i^r$'s are poorly determined in phenomenology
and have to be determined on the lattice.

Let us write the NNLO chiral expansion as 
\bea
   \ff(t)
   & = &
   \fflo + \ffnlo(t) + \ffnnlo(t) + \ffnnnlo(t)
   \hspace{3mm}
   (P\!=\!\pi^+,K^+,K^0),
   \label{eqn:chiral_fit:su3:ff}
   \\
   \ffnlo(t)
   & = & 
   \ffnloL(t) + \ffnloB(t),
   \hspace{5mm}
   \ffnnlo(t)
   = 
   \ffnnloL(t) + \ffnnloC(t) + \ffnnloB(t).
   \label{eqn:chiral_fit:su3:ff:contribu}
\eea
The terms $\fflo$, $\ffnlo$ and $\ffnnlo$ are 
the leading-order, NLO and NNLO contributions, respectively.
We add an even higher order correction $\ffnnnlo$, when necessary. 
The additional subscripts ``$L$'', ``$C$'' and ``$B$'' 
represent $L_i^r$-dependent, $C_i^r$-dependent and LEC-independent parts,
respectively.
The $C_i^r$-dependent NNLO analytic terms $\ffnnloC$ are given as~\cite{EMFF:ChPT:SU3}
\bea
   F_\pi^4\,
   \pffnnloC(t)
   & = & 
   -4 \Cppt\, M_\pi^2\, t  - 8 \Cpkt\, M_K^2\, t  - 4\Ctt\, t^2,
   \label{eqn:chiral_fit:su3:pff:nnlo-c}
   \\
   F_\pi^4\,
   \kpffnnloC(t)
   & = &
   -4 \Ckpt\, M_\pi^2\, t  - 4 \Ckkt\, M_K^2\, t  - 4\Ctt\, t^2,
   \label{eqn:chiral_fit:su3:kpff:nnlo-c}
   \\
   F_\pi^4\,
   \knffnnloC(t)
   & = &
  -\frac{8}{3}\, \Ckn\, (M_K^2 - M_\pi^2)\, t,
   \label{eqn:chiral_fit:su3:knff:nnlo-c}
\eea 
where the coefficients are linear combinations of $C_i^r$'s.
These are not independent:
$\Ckpt \!=\! \Cpkt + \Ckn/3$ and $\Ckkt \!=\! \Cppt + \Cpkt - \Ckn/3$.
We therefore treat the following four as fitting parameters
\bea
   \Cppt 
   & = & 
   4 C_{12}^r + 4 C_{13}^r + 2 C_{63}^r + C_{64}^r + C_{65}^r + 2 C_{90}^r, 
   \\
   \Cpkt
   & = &
   4 C_{13}^r + C_{64}^r,
   \hspace{5mm}
   \Ctt
   = 
   C_{88}^r - C_{90}^r, 
   \hspace{5mm} 
   \Ckn
   =
   2C_{63}^r - C_{65}^r.
\eea 

An example of the NNLO ChPT fit for the charged kaon EM form factor $\kpff$ 
is plotted 
in the left panel of Fig.~\ref{fig:em}. 
We obtain $\chi^2/\mbox{d.o.f}\!=\!1.8$ and
\bea
   L_9^r
   & = & 
   4.6(1.1)_{\rm stat}\left(^{+0.1}_{-0.5}\right)_{L_i^r} (0.4)_{a\!=\!0}\times 10^{-3},
   \hspace{5mm}
   \Ctt
   = 
   -6.4(1.1)_{\rm stat}(0.1)_{L_i^r}(0.5)_{a\!=\!0} \times 10^{-5},
   \hspace{10mm}
\eea
where the renormalization scale is set to $\mu\!=\!M_\rho$.
The second error is the uncertainty due to 
the choice of the input $L_{\{1,...,5\}}^r$, 
whereas the third is the discretization error estimated by power counting
$O((a \Lambda_{\rm QCD})^2)\!\sim\!8$\,\% with $\Lambda_{\rm QCD}\!=\!500$~MeV.
These results are reasonably consistent 
with the recent phenomenological estimate in Ref.~\cite{ChPT:LECs:SU2+SU3}. 
While other fit parameters are poorly known in phenomenology,
our results are roughly consistent with a naive power counting 
$C_i^r \!=\! O((4\pi)^{-4})$~\cite{EMFF:Nf3:JLQCD}.

An interesting observation in the right panel of Fig.~\ref{fig:em} is that
the non-trivial chiral correction $\kpffnlo+\kpffnnlo$ is largely dominated 
by the NLO analytic term $\kpffnloL$. 
Note that this term is not unexpectedly large, because our result for $L_9^r$ 
is consistent with the phenomenological estimate 
as well as a power counting $L_i^r \!=\! O((4\pi)^{-2})$.
Since this term is independent of the valence quark masses
($\pffnloL\!=\!\kpffnloL\!=\!2L_9^r t/F_\pi^2$),
the NNLO chiral expansion of the charged meson EM form factors $\pkpff$ 
shows reasonable convergence.
This is however not the case for the neutral kaon, 
because $\knffnloL$ vanishes to satisfy the constraint
$\knff(t)\!=\!0$ at $m_l\!=\!m_s$.

We obtain the following results for the charge radii from the NNLO ChPT fit
\bea
   \cradpff
   & = & 
   0.458(15)_{\rm stat}\left(^{+9}_{-1}\right)_{L_i^r}(37)_{a\ne 0}~\mbox{fm}^2,
   \hspace{3mm}
   \cradkpff
   = 
   0.380(12)_{\rm stat}\left(^{+7}_{-1}\right)_{L_i^r}(31)_{a\ne 0}~\mbox{fm}^2,
   \hspace{5mm}
   \\
   \cradknff
   & = &
   -0.055(10)_{\rm stat}(1)_{L_i^r}(4)_{a\ne 0}~\mbox{fm}^2.
\eea
These are in reasonable agreement with the experimental values
$\cradpff\!=\!0.452(11)~\mbox{fm}^2$, 
$\cradkpff\!=\!0.314(35)~\mbox{fm}^2$ and 
$\cradknff\!=\!-0.077(10)~\mbox{fm}^2$~\cite{PDG:2014}.


\section{Kaon semileptonic form factors}

\FIGURE{
   \label{fig:kl3:f0t_vs_Mpi2}
   \includegraphics[angle=0,width=0.48\linewidth,clip]%
                   {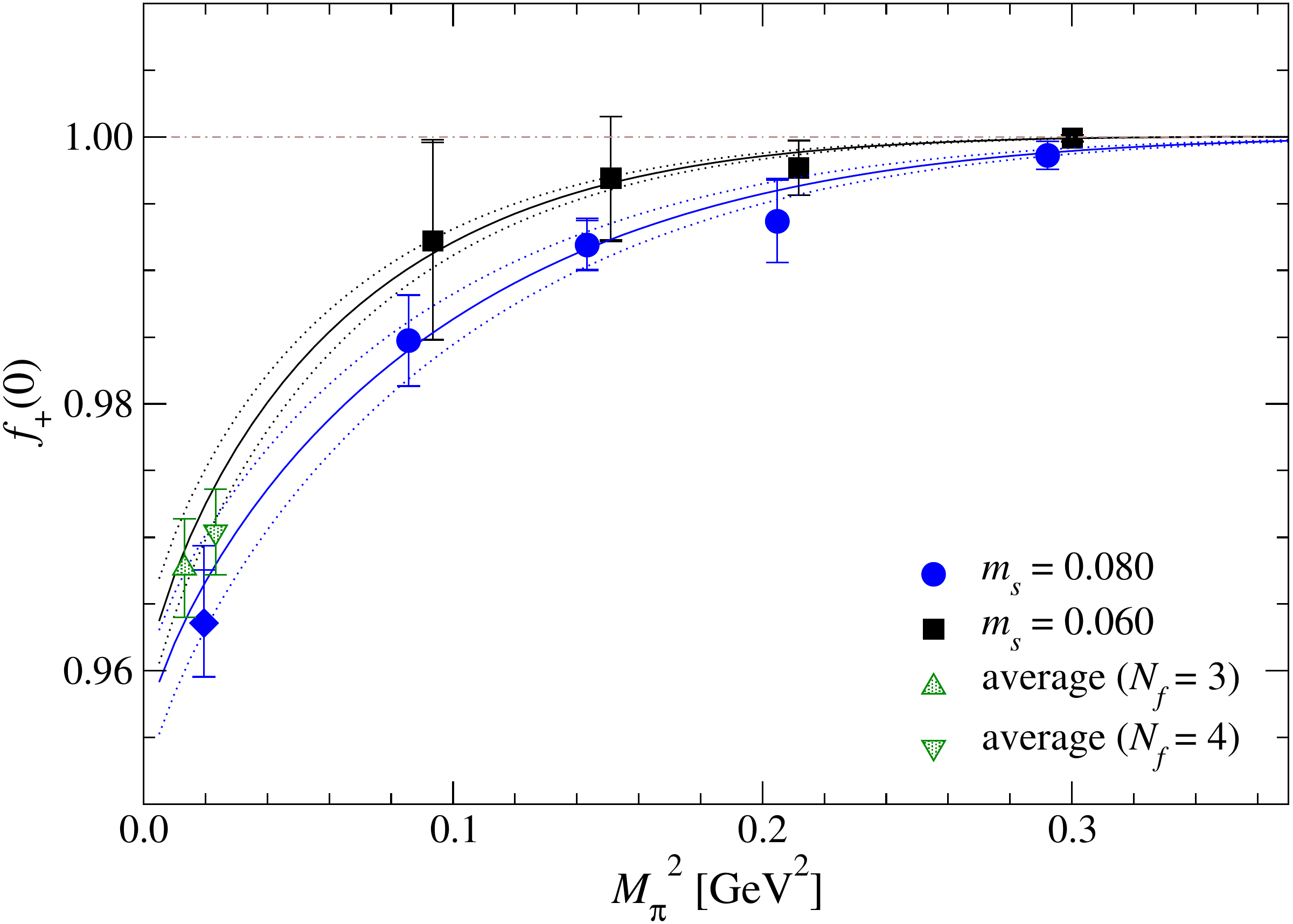}
   \vspace{-3mm}
   \caption{
      Extrapolation of $f_+(0)$ as a function of $M_\pi^2$ and $M_K^2$
      based on NNLO ChPT. Circles and squares show our data
      at $m_s\!=\!0.080$ and 0.060, respectively. 
      The diamond represents the value extrapolated to the physical point
      $(m_{l,\rm phys},m_{s,\rm phys})$.
      We also plot the preliminary FLAG averages for $N_f\!=\!3$ and 4
      by triangles.
   }
}
In Fig.~\ref{fig:kl3:f0t_vs_Mpi2},
we demonstrate a conventional determination of the normalization $f_+(0)$.
At simulated $M_\pi$'s,
we fix $f_+(0)$ 
assuming the following $t$ dependence of $f_+(t)$ and $f_0(t)$
\bea
   f_+(t) 
   & = & 
   f_+(0)
   \left\{
      \frac{1}{1-t/M_{K^*}^2} + a_+ t
   \right\},
   \hspace{5mm}
   \\
   f_0(t)
   & = & 
   f_+(0)\left( 1 + a_0 t + b_0 t^2 \right),
\eea
based on the vector meson dominance hypothesis.
The results are well described by the NNLO ChPT formula
with $\chi^2/{\rm d.o.f.}\!\sim\!0.2$.
The extrapolated value $f_+(0)\!=\!0.644(4)_{\rm stat}$ 
is consistent with the average of recent results~\cite{Kl3:Nf3:FNAL+MILC,Kl3:Nf4:FNAL+MILC,Kl3:Nf3:RBC/UKQCD}
by the Flavor Lattice Averaging Group (FLAG)~\cite{FLAG3,kaon:review:Lat15}.
We note that these recent studies are pursuing more precise determination 
by simulations near or directly at the reference point $t\!=\!0$ 
and the physical point $(m_{l,\rm phys},m_{s, \rm phys})$.
Note also that a study of the form factor shape based on 
a phenomenological parametrization
is also reported at this conference~\cite{Kl3:Nf4:ETM:Lat15}.
\clearpage

We employ a different strategy exploiting exact chiral symmetry:
we fit the lattice data of $f_+$ and $f_0$ to their NNLO ChPT formula
as a function of $M_{\{\pi,K\}}^2$ and $t$.
Similar to the EM form factors,
the chiral expansion of the vector form factor $f_+$~\cite{Kl3FF:ChPT:PS,Kl3FF:ChPT:BT}
has only $L_9^r$ at NLO and other $L_i^r$'s at NNLO. 
We fix $L_9^r$ to the value obtained in our analysis of the EM form factors,
whereas others are set to 
the phenomenological estimate~\cite{ChPT:LECs:SU2+SU3}.
The analysis of the EM form factors provides helpful information 
also for the NNLO LECs.
The coefficient $\Ctt$ in Eqs.~(\ref{eqn:chiral_fit:su3:pff:nnlo-c})\,--\,(\ref{eqn:chiral_fit:su3:kpff:nnlo-c}) also appears 
in the NNLO analytic term of $f_+$ as
\bea
   F_\pi^4 f_{+,4,C}
   & = & 
   \Cplspk\,(M_K^2-M_\pi^2)
  +c_{+,\pi t}^r\,M_\pi^2\,t
  +c_{+,K t}^r\,M_K^2\,t
  -4 \Ctt\, t^2.
\eea
Other two coefficients can be written 
in terms of those for the EM form factors 
\bea
  c_{+,\pi t}^r
  & = & 
  -4 \left( 
        2 C_{12}^r + 4 C_{13}^r + C_{64}^r + C_{65}^r + C_{90}^r
     \right)
  =
  -2 \left( \Cppt + \Cpkt - \Ckn \right),
  \\
  c_{+,K t}^r
  & = &
  -4 \left(
        2 C_{12}^r + 8 C_{13}^r + 2 C_{63}^r + 2C_{64}^r + C_{90}^r
     \right)
  =
  -2 \left( \Cppt + 3 \Cpkt + \Ckn \right).
\eea
The remaining one $\Cplspk\!=\!-8\left( C_{12}^r + C_{34}^r \right)$
describes SU(3) breaking effects at $t\!=\!0$, 
and hence is absent in the EM form factors.
Therefore we have only one free parameter $\Cplspk$ 
in the chiral extrapolation of $f_+$.

\begin{figure}[t]
\begin{center}
   \includegraphics[angle=0,width=0.48\linewidth,clip]%
                   {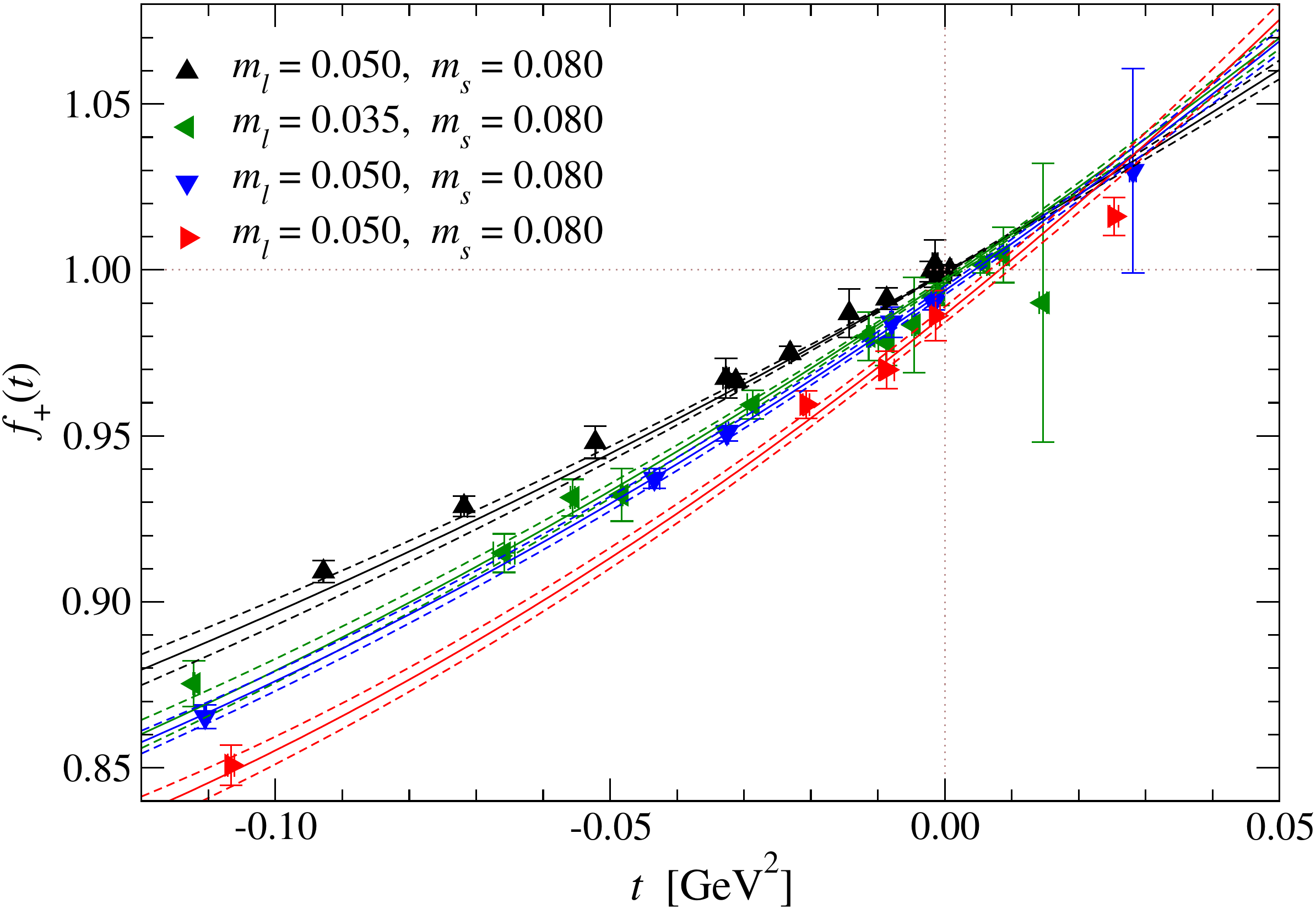}
   \hspace{3mm}
   \includegraphics[angle=0,width=0.48\linewidth,clip]%
                   {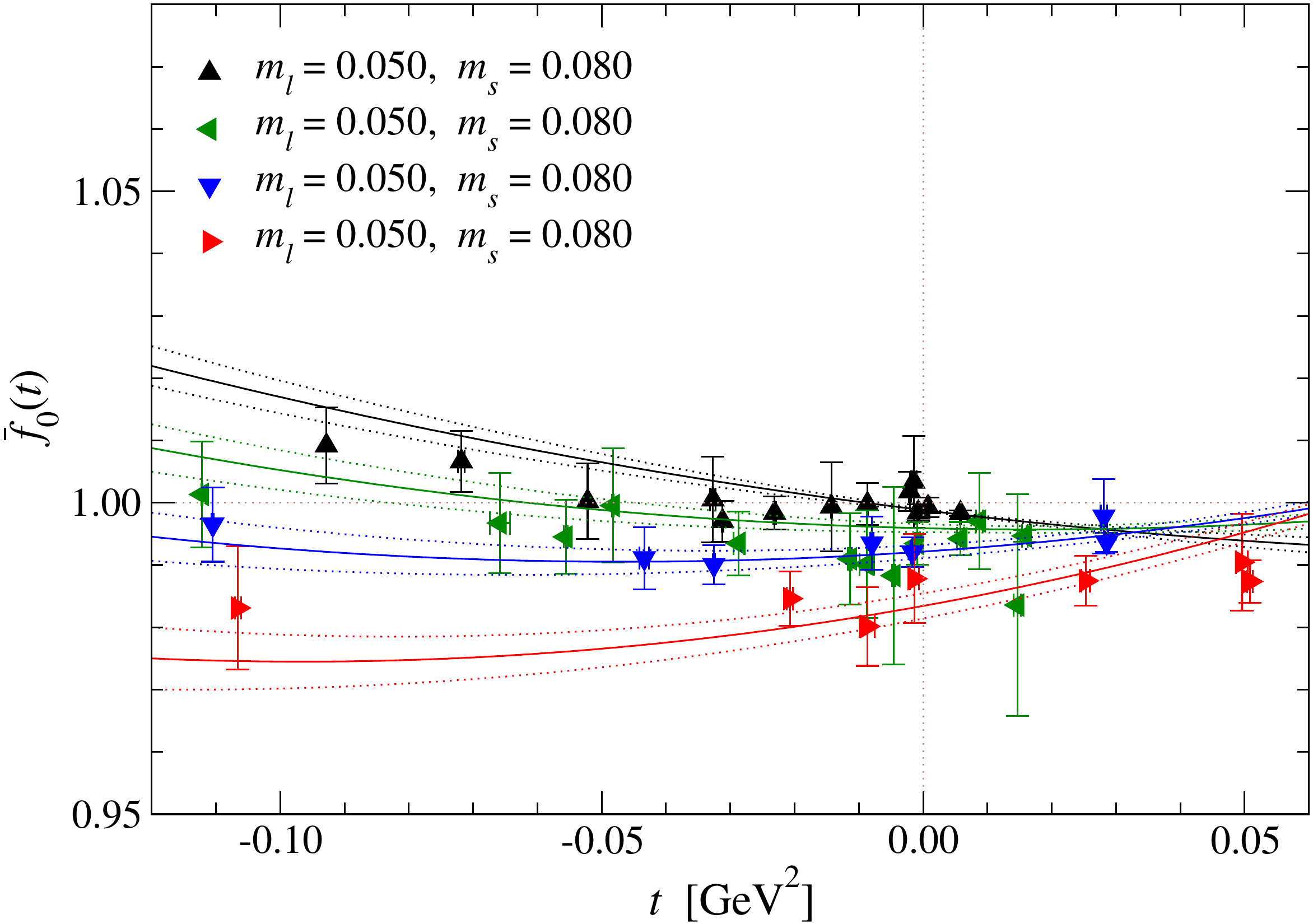}
   \vspace{-1mm}
   \caption{
     NNLO ChPT fit to $f_+(t)$ (left panel) and $\fzt(t)$ (right panel)
     as a function of $t$.
     Different symbols show data at different $M_\pi$'s.
   }
   \label{fig:kl3:f+_f0t_vs_t}
\end{center}
\vspace{-5mm}
\end{figure}

The scalar form factor $f_0$ has many additional NNLO LECs.
In order to carry out a chiral extrapolation with less free parameters,
we consider the following quantity proposed in Ref.~\cite{Kl3FF:ChPT:BT} 
\bea
   \tilde{f}_0(t)
   & = &
   f_0(t) + \frac{t}{M_K^2-M_\pi^2} \left( 1 - \frac{F_K}{F_\pi} \right).
\eea
The Dashen-Weinstein relation~\cite{DW_relation} suggests 
a large cancellation between the NNLO analytic terms
of $f_0$ and $F_K/F_\pi$.
In fact, the NNLO analytic term of $\fzt$ is given in a rather simple form 
\bea
   F_\pi^4   
   \tilde{f}_{0,4,C}(t)
   & = &
   \Cplspk \left( M_K^2 - M_\pi^2 \right)
  +\left( 8 C_{12}^r - \Cplspk \right) \left( M_K^2 + M_\pi^2 \right)\, t
 -8 C_{12} t^2,
\eea
and a simultaneous fit to $f_+$ and $\fzt$ has only two fitting parameters 
$\Cplspk$ and $C_{12}^r$.

Figure~\ref{fig:kl3:f+_f0t_vs_t} shows 
the NNLO ChPT fit to $f_+(t)$  and $\fzt(t)$ as a function of $t$.
Our data are well described 
with $\chi^2/{\rm d.o.f.}\!\sim\!0.7$. 
Similar to the EM form factors,
the $t$ dependence of $f_+$ is roughly approximated 
by the NLO analytic term $f_{+,2,L}\!=\!\kpffnloL$.
On the other hand, 
$\fzt$ has no NLO analytic term 
due to the cancellation between $f_0$ and $F_K/F_\pi$,
and shows a rather mild $t$ dependence.

From the simultaneous fit to $f_+$ and $\fzt$, we obtain 
\bea
  f_+(0)
  & = & 
  0.9636(36)_{\rm stat} \left(^{+41}_{-45}\right)_{\rm chiral} (29)_{a\ne0}
  = 
  0.9636\left(^{+62}_{-65}\right)
\eea
at the physical point. The first error is statistical.
The second is the systematic uncertainty of the chiral extrapolation,
which is estimated by repeating the fit including higher order corrections
or using different values for the input $L_{\{1,...,8\}}^r$. 
The third one is the discretization error estimated by power counting.
The total uncertainty is at the level of $\leq\!1$\,\%.

\FIGURE{
   \label{fig:kl3:lambda+p}
   \includegraphics[angle=0,width=0.48\linewidth,clip]%
                   {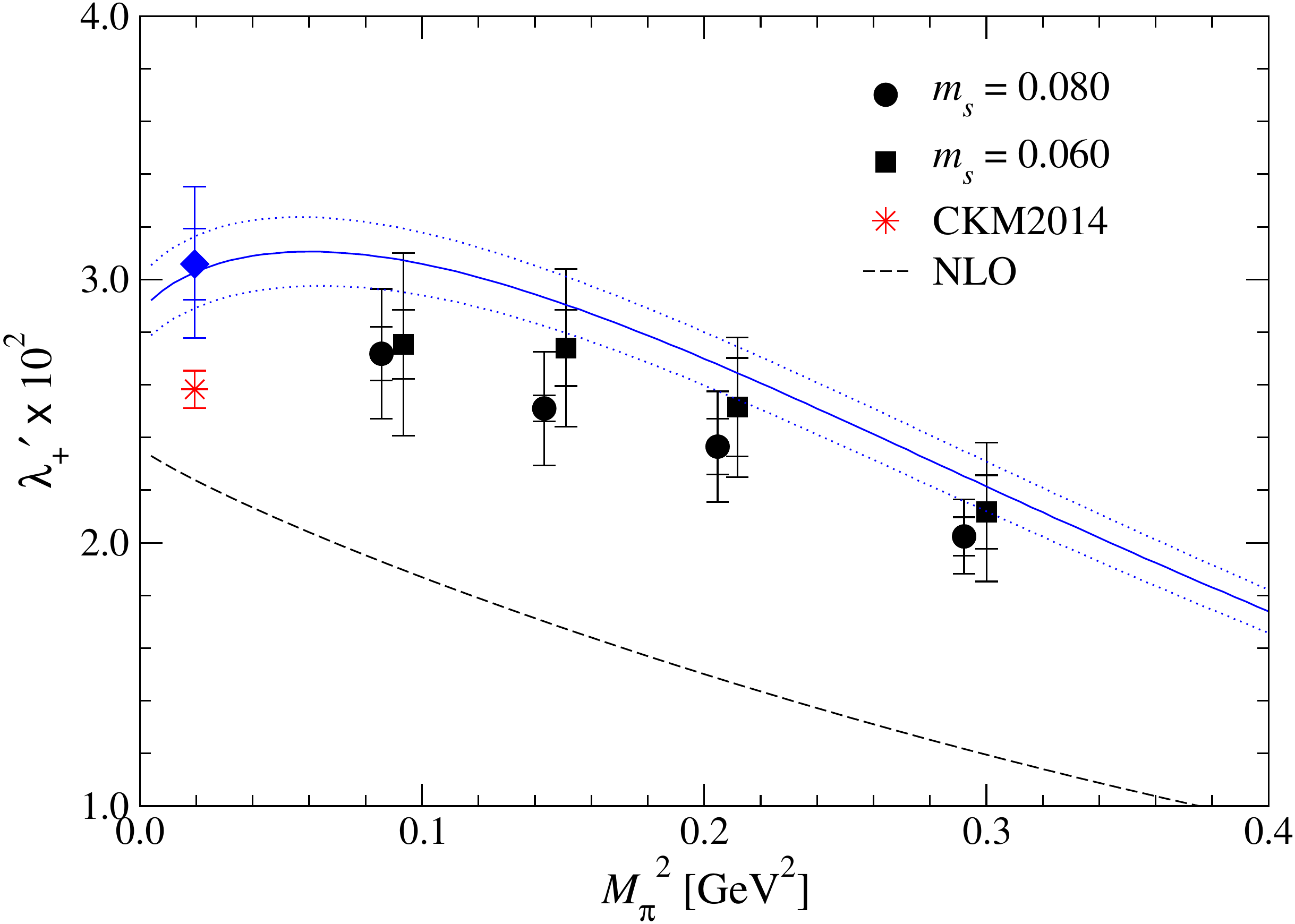}
   \vspace{-3mm}
   \caption{
      Slope $\lambda_+^\prime$ as a function of $M_\pi^2$.
      The solid line is reproduced from our NNLO ChPT fit,
      and the statistical error is shown by the dotted lines.
      The circles and squares represent 
      the values estimated at simulation points
      by assuming the polynomial parametrization (\ref{eqn:kl3:t-dep:f+}).
      The dashed line shows the NLO contribution.
   }
}
An advantage of our analysis method is that 
we can study both the normalization and shape of the form factors
by a unique fit based on NNLO ChPT.
In Fig.~\ref{fig:kl3:lambda+p},
we plot the slope $\lambda_+^\prime$ in the quadratic parametrization
\bea
   f_+(t)
   & = & 
   f_+(0)
   \left\{ 
      1 + \frac{\lambda_+^\prime}{M_{\pi^\pm}^2}t + O(t^2)
   \right\}.
   \hspace{10mm}
   \label{eqn:kl3:t-dep:f+}
   \eea
Namely,
\bea
   \lambda_+^\prime
   & = &
   \left.
      \frac{M_{\pi^\pm,\rm phys}^2}{f_+(0)} \frac{d f_+(t)}{dt}
   \right|_{t=0}.
   \label{eqn:kl3:t-dep:lambda+p}
\eea
The NNLO contribution turns out to be significant 
even near the physical point.
It is therefore important to study the form factor shape
by taking account of their non-analytic chiral behavior at NNLO.

From the NNLO ChPT fit, we obtain
$\lambda_+^\prime \!=\! 3.08(14)_{\rm stat}(31)_{\rm sys}\!\times\!10^{-2}$
and $\lambda_0^\prime\!=\!1.98(15)_{\rm stat}(44)_{\rm sys}\!\times\!10^{-2}$,
where we add the uncertainty of the chiral extrapolation
and the discretization error in quadrature.
This is consistent with recent experimental measurements,
$\lambda_+^\prime\!=\!2.58(7)\!\times\!10^{-2}$ and 
$\lambda_0^\prime\!=\!1.37(9)\!\times\!10^{-2}$~\cite{Kl3:Exp:CKM2014} 
within 2\,$\sigma$.
The largest uncertainty comes from the discretization for $\lambda_+^\prime$
and the chiral extrapolation for $\lambda_0^\prime$.


\section{Summary}

In this article, we have presented our lattice calculation
of the light meson EM and semileptonic form factors.
These form factors are precisely calculated
by using the all-to-all quark propagator.
Their chiral behavior is directory compared with continuum ChPT
by exploiting exact chiral symmetry preserved with the overlap quark action.

We observe that the lattice data of the charged meson EM form factors 
are reasonably well described by their NNLO ChPT formula,
and estimate the relevant NLO and NNLO LECs. 
These results are used in the chiral extrapolation 
for the kaon semileptonic form factors, 
and their normalization $f_+(0)$ is determined with sub-\% accuracy.
We confirm reasonable consistency of the form factor shape,
namely the charge radii and slopes of the semileptonic form factors,
with experiment.

One of the largest uncertainty is the discretization error at a finite lattice 
spacing. It is important to extend this study to finer lattices 
and heavy flavor physics. 
Simulations in these directions are underway~\cite{Noaki:Lat14}
using a computationally cheaper fermion action 
with good chiral symmetry~\cite{Kaneko:Lat13}.

\vspace{3mm}

We thank Johan Bijnens for making his code 
to calculate pion and kaon form factors in NNLO ChPT available to us.
Numerical simulations are performed on Hitachi SR16000 and 
IBM System Blue Gene Solution at KEK
under a support of its Large Scale Simulation Program (No.~15/16-09),
and on SR16000 at YITP in Kyoto University.
This work is supported in part by the Grant-in-Aid of the MEXT
(No.~25287046, 26247043, 26400259 and 15K05065)
and by MEXT SPIRE and JICFuS.


\begin{thebibliography}{99}

\bibitem{exW}
H.~Fukaya {\it et al.} (JLQCD Collaboration),
Phys. Rev. D {\bf 74}, 094505 (2006)
[arXiv:hep-lat/0607020].

\bibitem{fixedQ}
S.~Aoki, H.~Fukaya, S.~Hashimoto and T.~Onogi,
Phys. Rev. D {\bf 76}, 054508 (2007)
[arXiv:0707.0396 [hep-lat]].

\bibitem{Kl3:Nf3:JLQCD:Lat12}
T.~Kaneko {\it et al.} (JLQCD Collaboration),
PoS {\bf Lattice 2012} 111 (2012)
[arXiv:1211.6180 [hep-lat]].

\bibitem{EMFF:Nf3:JLQCD}
S.~Aoki {\it et al.} (JLQCD Collaboration),
arXiv:1510.06470 [hep-lat].

\bibitem{ChPT:LECs:SU2+SU3}
J.~Bijnens and G.~Ecker, 
Ann. Rev. Nucl. Part. Sci. {\bf 64}, 149 (2014)
[arXiv:1405.6488 [hep-ph]].

\bibitem{EMFF:ChPT:SU3}
J.~Bijnens and P.~Talavera,
JHEP {\bf 0203}, 046 (2002)
[arXiv:hep-ph/0203049].

\bibitem{PDG:2014}
K.A.~Olive {\it et al.} (Particle Data Group), 
Chin. Phys. C, 38, 090001 (2014).

\bibitem{Kl3:Nf3:FNAL+MILC}
A.~Bazavov {\it et al.} (Fermilab Lattice and MILC Collaborations),
Phys. Rev. D {\bf 87}, 073012 (2013)
[arXiv:1212.4993 [hep-lat]].


\bibitem{Kl3:Nf4:FNAL+MILC}
A.~Bazavov {\it et al.} (Fermilab Lattice and MILC Collaborations),
Phys. Rev. Lett. {\bf 112}, 112001 (2014)
[arXiv:1312.1228 [hep-ph]].


\bibitem{Kl3:Nf3:RBC/UKQCD}
P.A.~Boyle {\it et al.} (RBC/UKQCD Collaboration),
arXiv:1504.01692 [hep-lat].

\bibitem{FLAG3}
S.~Aoki {\it et al.} (Flavor Lattice Averaging Group),
in preparation.

\bibitem{kaon:review:Lat15}
A.~J\"uttner, in these proceedings. 

\bibitem{Kl3:Nf4:ETM:Lat15}
L.~Riggio {\it et al.} (ETM Collaboration),
in these proceedings.

\bibitem{Kl3FF:ChPT:PS}
P.~Post and K.~Schilcher,
Eur. Phys. J. C {\bf 25}, 427 (2002)
[arXiv:hep-ph/0112352].

\bibitem{Kl3FF:ChPT:BT}
J.~Bijnens and P.Talavera,
Nucl. Phys. B {\bf 669}, 341 (2003)
[arXiv:hep-ph/0303103].

\bibitem{DW_relation}
R.F.~Dashen, L.~Ling-Fong, H.~Pagels and M.~Weinstein,
Phys. Rev. D {\bf 6}, 834 (1972).

\bibitem{Kl3:Exp:CKM2014}
M.~Moulson, 
arXiv:1411.5252 [hep-ex].

\bibitem{Noaki:Lat14}
J.~Noaki {\it et al.} (JLQCD Collaboration),
PoS {\bf LATTICE2014}, 069 (2015).

\bibitem{Kaneko:Lat13}
T.~Kaneko {\it et al.} (JLQCD Collaboration),
PoS {\bf LATTICE2013}, 125 (2014)
[arXiv:1311.6941 [hep-lat]].

\end{thebibliography}
\end{document}